\documentclass[a4paper,11pt]{article}
\usepackage{jcappub} % for details on the use of the package, please see the JINST-author-manual
\arxivnumber{2306.17488} % Only if you have one
\title{Mass and tidal parameter extraction from gravitational waves of binary neutron stars mergers using deep learning}

\author[a,b,c]{Shriya Soma}
\author[a,b,d]{Horst St\"ocker}
\author[a,e, 1]{Kai Zhou\note{Corresponding author.}}

\affiliation[a]{Frankfurt Institute for Advanced Studies (FIAS), D-60438 Frankfurt am Main, Germany}
\affiliation[b]{Institut f\"ur Theoretische Physik, Goethe Universit\"at, D-60438 Frankfurt am Main, Germany}
\affiliation[c]{Xidian-FIAS International Joint Research Center, D-60438 Frankfurt am Main, Germany}
\affiliation[d]{GSI Helmholtzzentrum f\"ur Schwerionenforschung GmbH, D-64291 Darmstadt, Germany}
\affiliation[e]{School of Science and Engineering, The Chinese University of Hong Kong, Shenzhen 518172, China}

\emailAdd{zhou@fias.uni-frankfurt.de}

\abstract{Gravitational Waves (GWs) from coalescing binaries carry crucial information about their component sources, like mass, spin and tidal effects. This implies that the analysis of GW signals from binary neutron star mergers can offer unique opportunities to extract information about the tidal properties of NSs, thereby adding constraints to the NS equation of state. In this work, we use Deep Learning (DL) techniques to overcome the computational challenges confronted in conventional methods of matched-filtering and Bayesian analyses for signal-detection and parameter-estimation. We devise a DL approach to classify GW signals from binary black hole and binary neutron star mergers. We further employ DL to analyze simulated GWs from binary neutron star merger events for parameter estimation, in particular, the regression of mass and tidal deformability of the component objects. 
The results presented in this work demonstrate the promising potential of DL techniques in GW analysis, paving the way for further advancement in this rapidly evolving field. The proposed approach is an efficient alternative to explore the wealth of information contained within GW signals of binary neutron star mergers, which can further help constrain the NS EoS. 
}


\begin{document}
\maketitle
\flushbottom

\section{Introduction}
The era of gravitational wave astronomy commenced with LIGO's first detection of gravitational waves~(GWs) from the collision of two black holes on 14$^{\text{th}}$ September 2015~\cite{LIGOScientific:2016emj}. 
Since then, the LIGO-Virgo Scientific collaboration has made several GW detections from compact binary coalescences; 11 events in the first and second observing runs~(O1 and O2), and 79 in the third observing run~(O3)~\cite{LIGOScientific:2018mvr, LIGOScientific:2020ibl, LIGOScientific:2021usb, LIGOScientific:2021djp}. 
These events comprise mergers of binary black holes~(BBHs), binary neutron stars~(BNSs), neutron star-black hole~(NSBH) binaries~\cite{LIGOScientific:2021qlt} and also component objects from the `mass gap'~\cite{LIGOScientific:2020zkf}. GW170817, the first GW event from a BNS merger detected by Advanced LIGO and Virgo, marked a major advancement in the ongoing research on neutron stars (NSs)~\cite{LIGOScientific:2017vwq}. 

Prior to the event GW170817, the NS equation of state (EoS) in the intermediate density range~(2-7{\it n}$_s$, where {\it n}$_s$ is the nuclear saturation density) was mainly constrained by precise mass measurements of pulsars, i.e., any EoS that does not satisfy the minimum lower band set by the highest neutron star mass observed (currently, PSR~J0952\ensuremath{-}0607 with a pulsar mass measurement of 2.35$\pm$0.17$M_{\odot}$~\cite{Romani:2022jhd}, followed by more precise measurements of PSR~J1810+1744 (2.13$\pm$0.04$M_{\odot}$~\cite{Romani:2021xmb}), PSR~0740+6620 (2.08$\pm$0.07$M_{\odot}$~\cite{Fonseca:2021wxt}), PSR~J0348+0432 (2.01$\pm$0.04M$_{\odot}$~\cite{Antoniadis:2013pzd})), can be eliminated. With the advancement of the GW detectors, it is now possible to also extract the 
tidal deformability~($\Lambda$), another parameter that constrains the EoS.
The tidal effects in a BNS merger are reflected in the GW phase evolution at late inspirals of the event~\cite{Hinderer:2009ca}. GW170817 has allowed for the very first time the extraction of the tidal parameter~\cite{LIGOScientific:2018hze}, setting additional constraints on the EoS. The dependence of tidal deformability on radius has made it possible to estimate the radius of a canonical 1.4$M_{\odot}$ neutron star~\cite{Raithel:2018ncd, Zhao:2018nyf, Soma:2019utv}. These estimates provide complementary information to the independent radii measurements by the NICER~(Neutron Star Interior Composition Explorer) collaboration.
NICER reported its first results on radius measurements of the pulsar, PSR~J0030+0451~\cite{Miller:2019cac, Riley:2019yda} in 2019, followed by the radius measurements of PSR~0740+6620~\cite{Miller:2021qha, Riley:2021pdl} in 2021. The combined results from NICER and GW170817 allow for constraints on the EoS, and estimates for the radius and moment of inertia of a 1.4$M_{\odot}$ NS~\cite{Jiang:2019rcw}. These estimates are excellent probes for dense matter in the relatively low density range~({\it n}$_s$ - 3{\it n}$_s$)~\cite{Lattimer:2019eez}. In addition, different studies of the electromagnetic observations following GW170817, provide estimates on the upper limit of the maximum neutron star mass \cite{Rezzolla:2017aly, Shibata:2019ctb}. Furthermore, a new lower bound on the maximum mass was obtained from the analysis of the event GW190814, based on the assumption that the secondary mass was a rapidly spinning neutron star~\cite{Most:2020bba}. These results could help constrain the EoS in the high density regime~(3{\it n}$_s$ - 5{\it n}$_s$). All these findings demonstrate that gravitational waves are instrumental in providing direct or indirect constraints on the dense matter EoS.

Gravitational waveforms can be approximated using the post-Newtonian~(PN) formalism. In the PN expansion, the chirp mass,~$\mathcal{M}$, is fairly easy to determine from the leading order term~\cite{Chatziioannou:2020pqz}~(i.e., given the frequency and its time derivative). The mass ratio appears only in higher order terms and is harder to estimate due to degeneracy with the aligned spin components~\cite{Cutler:1994ys}. The combined tidal deformation,~$\Tilde{\Lambda}$, on the other hand, appears only at the 5PN order~\cite{Flanagan:2007ix, Chatziioannou:2020pqz, Wade:2014vqa}. These tidal effects show up at the high frequency orbits of the binary. 
At such frequencies, the detector noise is comparatively large and as a result, the recovery of tidal deformability is significantly affected~\cite{Wade:2014vqa}. Despite the challenges faced in retrieving tidal properties from BNS mergers, ample information on the underlying EoS can be obtained from its extraction. Therefore, the estimation of tidal deformability from gravitational waves has drawn considerable attention over the years~\cite{Chatziioannou:2020pqz}. 
A universal relation between tidal deformability and the GW frequency at peak amplitude was established for equal mass binaries~\cite{Read:2013zra}. This implies that regardless of the EoS, one can estimate the tidal deformability, given the GW frequency at peak amplitude. A similar relation was discovered between tidal deformability and the dominant post-merger GW frequency for hadronic EoSs~\cite{Bernuzzi:2015rla, Rezzolla:2016nxn, Bauswein:2018bma, Lioutas:2021jbl}. However, the universality is violated when a phase transition occurs from hadronic to deconfined quark matter~\cite{Bauswein:2018bma}. Therefore, independent measurements of the tidal parameter from GW signals are important to understand the dense matter EoS.
With the continuous development of detectors, many more GW observations are anticipated in the coming years. Future BNS merger events, as a result of an increase in the detector sensitivities, are expected to enhance our knowledge of dense matter by constraining the EoS further~\cite{ KAGRA:2013rdx, Iacovelli:2022bbs}. 

A recent alternative approach to probe the dense matter EoS is through Deep Learning~(DL). It has been successfully applied to explore the properties of NS matter~\cite{Fujimoto:2021zas, Morawski:2020izm, Soma:2022vbb, soma:2022qnv, Krastev:2021reh, Farrell:2022lfd, Carvalho:2023ele, Krastev:2023fnh, Chatterjee:2023ecc} as well as matter in the regime of perturbative quantum chromodynamics~(pQCD)~\cite{Pang:2016vdc, Jiang:2021gsw, Zhou:2023pti}. 
The demand for efficient handling of data from GW detectors for signal-detection and parameter estimation makes DL techniques particularly promising in the field of GW astronomy~\cite{George:2016hay, Krastev:2019koe, Shen:2019ohi, Morawski:2019awi, Verma:2021epx, Jadhav:2023mqx}. More specifically, it has been shown that DL is very reliable for the real-time GW detection of BBH mergers and subsequent parameter estimation~\cite{George:2016hay, Green:2020hst, Dax:2021tsq, Dax:2022pxd}, rapid identification of transient GW signals from BNS mergers~\cite{Krastev:2019koe}, forecasting BNS mergers~\cite{Wei:2020sfz}, denoising GWs~\cite{Shen:2019ohi}, identifying glitches in GW data~\cite{Mesuga:2021qeq}, and for the detection and classification of NSBH mergers~\cite{Qiu:2022wub}. 
So far, DL methods for GW analysis have focused largely on signals from binary black hole mergers. Although deep learning has also been used in the GW analysis of binary neutron star mergers~\cite{Krastev:2019koe, Wei:2020sfz, Krastev:2020skk}, there has been little attention on the extraction of tidal deformability. In this article, we present deep learning methods to analyze gravitational waves for both signal-detection (a classification task) and parameter-estimation (a regression task). In the first part of the article, we discuss the detection of simulated GWs originating from coalescing BBHs and BNSs. In the second part of the work, we focus on the estimation of two source parameters from GWs of binary neutron star mergers, namely, the chirp mass and dimensionless combined tidal deformability. We further illustrate that the extraction of both $\mathcal{M}$ and $\Tilde{\Lambda}$ could provide direct constraints on the NS EoS.

The article is organized as follows: (i) The data preparation methods for both classification and regression are presented in section~\ref{sec:data_prep}. (ii) In section~\ref{sec:Classification}, we describe the GW signal-detection from BBH and BNS mergers. The section entails the DL methods used in the analysis as well as results obtained from the trained DL model. (iii) In section~\ref{sec:Regression}, we detail the estimation of chirp mass and combined tidal deformability from BNS mergers. We present the neural network architecture for the regression task and the results obtained from the various trained NNs. (iv) We end with discussions and conclusions in section~\ref{sec:discuss}.

\section{Data Preparation} \label{sec:data_prep}
In this section, we discuss the methods of data generation for GW analysis. 
In order to analyze data for GW signal-detection and parameter-estimation, the standard conventional method is the template-matched filtering~\cite{DalCanton:2014hxh}. This approach involves the use of template banks that contain waveforms from a large set of parameter space, making the process computationally expensive. In order to circumvent this challenge, we introduce neural networks~(NNs) for analyzing the simulated gravitational waveforms. NNs prove advantageous for such tasks as the computational challenge only occurs during the training process. The training process, however, requires large amounts of data, which we generate using waveform-approximant models. Once trained, the neural network can be used to perform a prompt analysis on the test data.

Since Numerical Relativity (NR) simulations are computationally very expensive for generating accurate BNS waveforms, we use GW approximants from the LALSuite library~\cite{lalsuite}. Model waveforms like the PhenomDNRT, PhenomPNRT, SEOBNRT, etc, which use the aligned-spin point-particle model (with and without precession effects), and the aligned-spin point-particle effective-one-body (EOB) model~\cite{Santamaria:2010yb, Husa:2015iqa, Khan:2015jqa, Hannam:2013oca, Bernuzzi:2014owa, Hinderer:2016eia, Purrer:2015tud}, perform up to a high level of accuracy when tested with NR simulations. These waveforms are generated by adding a tidal amplitude correction to their corresponding BBH baseline waveforms. In particular, we use the inspiral-merger-ringdown~(IMR) precession model ``IMRPhenomPv2\_NRTidalv2"~\cite{Dietrich:2019kaq} for simulating gravitational waveforms of BNS mergers. The BBH merger waveforms are modeled with the same baseline waveform, ``IMRPhenomPv2" approximant, i.e., without tidal effects. 
Since the phenomenological waveform models are typically developed and constructed in the frequency-domain, we generate the BBH and BNS merger simulations in the frequency series~(in contrast to Ref.~\cite{Krastev:2019koe}, where the waveforms were analyzed in the time-domain). All waveforms are generated without spin.

\begin{figure}[!htbp]
    \centering
    \includegraphics{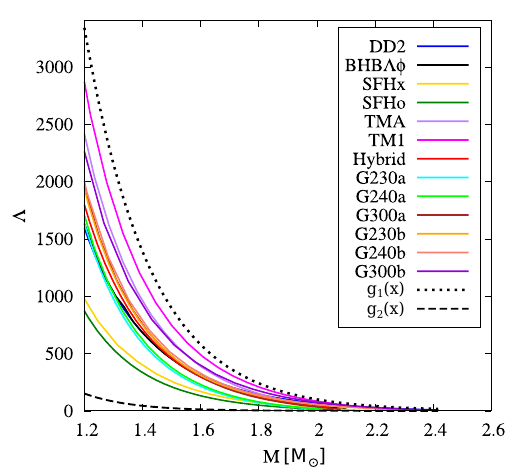}
    \caption{$\text{M}-\Lambda$ curves of several EoSs taken from~\cite{Soma:2019utv} are shown as solid colored curves. The black dotted and dashed curves labeled as $g_1(x)$ and $g_2(x)$, respectively, are chosen functions to span a wide range in the $\text{M}-\Lambda$ space. The $\text{M}-\Lambda$ data points enveloped in this region are used to simulate GW signals of BNS mergers in the regression task.}
    \label{td_m_fits}
\end{figure}

The BBH merger components are chosen to have masses within the range~[5,~50]$M_{\odot}$. The masses are selected randomly assuming a uniform distribution in the specified mass range. 
The BNS systems are considered to have component masses in the range~[1.2,~2.2]$M_{\odot}$, such that the higher mass is labeled as the primary mass, $m_1$, and lower mass as the secondary mass, $m_2$. 
For the BNS signals, apart from the masses, we also input the tidal deformabilities of the component NSs. We do not rely on a specific EoS for generating the tidal deformabilities, i.e.,~we generalize the $\text{M}-\Lambda$ relation. 
In order to define a large area in the $\text{M}-\Lambda$ space, we use two fit functions, $g_1(x) = a_1\exp(b_1x)$ and $g_2(x) = a_2\exp(b_2x)$ as upper and lower envelopes to several microphysical EoS models as shown in figure~\ref{td_m_fits}. The tidal deformability for each star is assigned a random value from the range lying within the envelope marked by the dotted~($g_1(x)$) and dashed~($g_2(x)$) black lines in figure~\ref{td_m_fits}. Once a primary mass,~$m_1$, is chosen, the corresponding tidal deformability, $\Lambda_1$, is randomly assigned a value that lies between the two fit functions $g_1(m_1)$ and $g_2(m_1)$. Here, we use $a_1$~=~6.45e+05, $b_1$~=~-4.386, $a_2$~=~2.45e+05, and $b_2$~=~-6.16. The tidal deformability, $\Lambda_2$, of the secondary mass~($m_2$), is then chosen from a uniform distribution of values lying within the range ($\Lambda_1$,~$g_1(m_2)$]. This way we ensure that $\Lambda$ is a monotonically decreasing function of $\text{M}$ in the region of interest~\footnote{In this approach, we ignore the possibility of stable twin stars, where the tidal deformability depicts two distinct branches for the same mass NS~\cite{Montana:2018bkb}.}. 
Note that the EoSs in figure~\ref{td_m_fits} are not all necessarily within the $\Lambda$-constraints set by GW170817 as shown in~\cite{Soma:2019utv}. However, this choice of an envelope ensures that the $\text{M}-\Lambda$ range of relevance is enclosed by the parameters we use for the simulations.

%%%%%%%%%%%%%%%%%%%%%%%%%%%%%%%%%%%%%%%%%%%%%%%%%%%%%%%%
%%%%%%%%%%%%%%%%%%%%%%%%%%%%%%%%%%%%%%%%%%%%%%%%%%%%%%%%

\begin{figure}[!htbp]
    \centering
    \includegraphics[width=0.5\columnwidth]{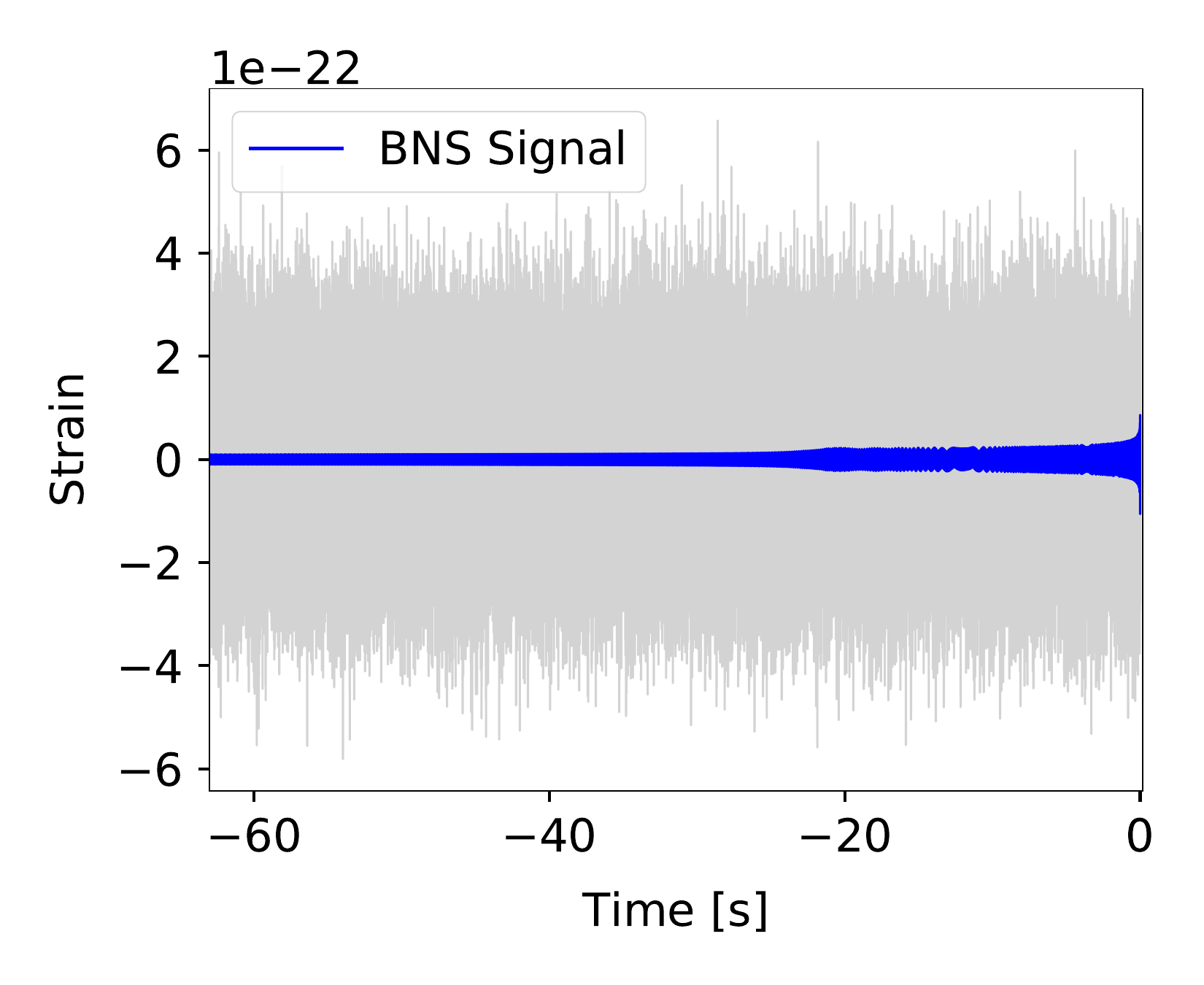}
    \caption{Example of a time-series BNS merger signal injected into aLIGO's colored noise~(grey). The blue waveform represents a clean signal from the inspiral and merger of an asymmetric-mass BNS system ($m_1=1.75M_{\odot}, m_2=1.50M_{\odot}$ and $\Lambda_1=176.94, \Lambda_2=450.05$, based on the DD2 EoS). This example showcases a BNS signal with oSNR=20. }
    \label{fig:ns_osnr20}
\end{figure}

All waveforms are generated within the frequency range, $f\in$ [$f_{\text{low}}$,~$f_{\text{high}}$]~Hz. 
Here, we employ the conventional methods of noise generation in GW astronomy. We generate a colored noise background from the ``Advanced LIGO Zero-Detuned High Power" power-spectral density (PSD). The simulated BBH and BNS inspiral, merger signals are injected into this noise background to attain optimal signal-to-noise ratio (oSNR) values lying in the range,~$20-30$. The optimal SNR ($\rho _{opt}$) is defined as
\begin{equation}
\rho _{opt} = 2 \left[ \int_0^{\infty} df \frac{\mid h(f) \mid ^2}{S_n (f)}\right] ^{1/2}
\end{equation}
where $h(f)$ is the gravitational wave signal in the frequency domain, and $S_n(f)$ is the detector PSD~\cite{Sathyaprakash:2009xs}. These signals are whitened for further analysis in the classification and regression networks. In figure~\ref{fig:ns_osnr20}, we present an example of a simulated waveform in the time-domain, where a BNS inspiral-merger signal is injected into detector noise. The grey waveform includes the injected signal in the colored noise. The clean inspiral-merger signal of an asymmetric-mass BNS system~($m_1=1.75M_{\odot}, m_2=1.50M_{\odot}$ and $\Lambda_1=176.94, \Lambda_2=450.05$, based on the DD2 EoS~\cite{Typel:2009sy}) is represented in blue, and has an oSNR value of 20. Additionally, we also perform analyses on clean signals, i.e., simulated BBH or BNS merger waveforms without any noise. We generate 75,000 such waveforms belonging to each category. 

%%%%%%%%%%%%%%%%%%%%%%%%%%%%%%%%%%%%%%%%%%%%%%%%%%%%%%%%
%%%%%%%%%%%%%%%%%%%%%%%%%%%%%%%%%%%%%%%%%%%%%%%%%%%%%%%%

\section{Classification of GW signals from different sources} \label{sec:Classification}
In this section, we provide details on the architecture of the classification network~(subsection~\ref{subsec:dl_classification}) as well as the associated results~(subsection~\ref{subsec:results_classification}).

\subsection{Classification: Network Architecture} \label{subsec:dl_classification}
The large amount of data associated with GW data calls for efficient methods of data analysis. 
Here, we discuss the use of deep neural networks for the classification of GW signals from different sources, specifically, BBH and BNS mergers.
We generate the GW signals of binary black hole and binary neutron star mergers for classification using the methods described in section~\ref{sec:data_prep}. The network designed for classification consists of the input layer, several 1D~convolutional layers and a couple of dense layers. The output layer of the network contains three nodes for classifying the input waveform into one of three categories: BBH mergers, BNS mergers, or no signals. The structure of the classification network is detailed in table~\ref{cla_cnn}. 
We choose $f_{\text{low}}$~=~23~Hz, $f_{\text{high}}$~=~1024~Hz and $\Delta f$~=~1/64. Therefore, the length of the input waveform is 64063. The convolution layers have 32 or 64 filters as mentioned in the table. In addition, the layers use kernels of size 16 and strides of 4, 4, 2, 2, and 2 in that respective order. The kernel weights are initialized from the {\it{He}}~Normal distribution~\cite{he2015delving}. The leaky rectified linear unit or `LeakyReLU' non-linear activation~($\alpha = 0.05$) is applied to all the hidden layers. 
The `softmax' function is applied to the last layer. The target labels are one hot encoded.
A waveform in the frequency domain,~$h(f)$, is written as $A(f)\exp{(i\phi(f))}$ where $A$ is the Fourier amplitude or absolute value and $\phi$ is the phase or argument. The absolute and phase values of the gravitational waveforms are used as input for the classification network. The utilization of two input channels to the network results in the shape of the network (64063,~2). The first channel, or the Fourier amplitude is first normalized to have a unit integral value. The waveforms are then scaled~(with the maximum values) to lie within the range~[0,~1]. The second dimension, with the phase,~$\phi$, is scaled with $\pi$, such that it lies within~[0,~1]. 
We use the cross entropy loss function and an Adam optimizer~\cite{2014arXiv1412.6980K} with a learning rate of 10$^{-4}$ for the classification network. 
From all the generated signals, 66,000 waveforms from each category, i.e., a total of 198,000 waveforms from the three categories are shuffled and used to train the classification network. A batch size of 16 is employed to fit the network model to the training data. The validation data uses 9,000 waveforms from each category, adding up to a total of 27,000 waveforms. 
The training is initiated with the clean signals and then continued on signals with lower SNR values (30, 25, and 20). The network is trained for 15 epochs in the case of clean signals, and for 30 epochs in the case of SNR=30. However, we increase the number of epochs as we decrease the SNR value. For SNR values of 25 and 20, we train the network for 60 and 100 epochs, respectively. The results of the classification network for different pSNR values are discussed in the next subsection,~\ref{subsec:results_classification}. 
%%%%%%%%%%%%%%%%%%%%%%%%%%%%%%%%%%%%%%%%%%%%%%%%%%%%%%
\begin{table}[ht!]
\caption{Classification model architecture used for classifying the simulated input GW signals into 3 classes, i.e., BBH mergers, BNS mergers or noise. The network consists of 2,094,787 trainable parameters.}
\centering
\def\arraystretch{1.4}
\setlength\tabcolsep{12pt}
\begin{tabular}{@{}ccc@{}}
\hline\hline
Layer Index & Layer & Dimension\\
\hline%
1 & Input & 64063 x 2\\
%2 & Reshape & 128257 x 1\\
2 & Convolution 1D & 16012 x 32\\
3 & Convolution 1D & 4000 x 32\\
4 & Convolution 1D & 1993 x 64\\
5 & Convolution 1D & 989 x 32\\
6 & Convolution 1D & 487 x 32\\
- & Flatten & 15584\\
7 & Dense Layer & 128\\
- & Dropout & 128\\
8 & Dense Layer & 3\\
- & Output & 3\\
\hline\hline
\end{tabular}
\label{cla_cnn}
\end{table}
%%%%%%%%%%%%%%%%%%%%%%%%%%%%%%%%%%%%%%%%%%%%%%%%%%%%%%
\subsection{Classification: Results}\label{subsec:results_classification}
\begin{figure*}[!htbp]
    \centering
    \includegraphics[width=0.94\textwidth]{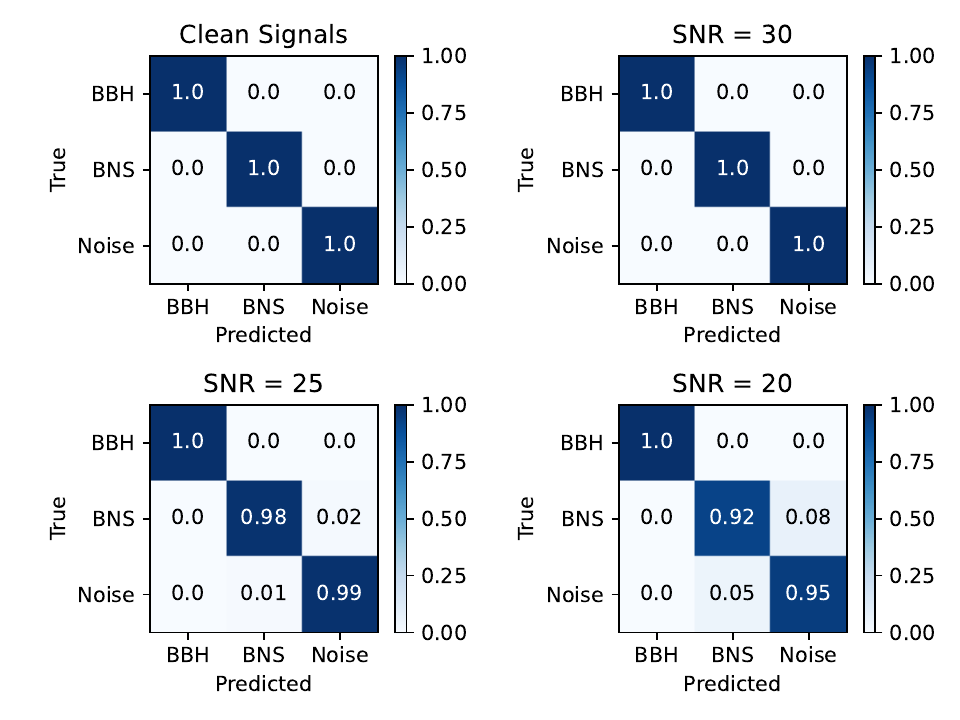}
    \caption{Normalized confusion matrices of the classification network for different values of SNR. The networks are trained for 15, 30, 60 and 100 epochs in case of clean signals, and SNR values of 30, 25 and 20, respectively. The number of misclassifications increases with decreasing SNR. We obtain macro-average precision values of 1.0 for clean signals; $\sim$1.0, 0.99 and 0.96 for SNR of 30, 25 and 20, respectively.}
    \label{fig:classify_conf_matrix}
\end{figure*}

We discuss the results obtained from the neural network for classifying simulated GW signals from different sources. The trained classification network predicts the class that a test signal belongs to, i.e., one of binary black holes mergers, binary neutron star mergers, and a third class to detect waveforms that only include noise. These predictions are compared against the true labels in the form of a confusion matrix. The confusion matrices provide details such as the number of true positives~(TP), true negatives~(TN), false positives~(FP) and false negatives~(FN), which are used to measure the performance of the trained model. We can quantify the performance utilizing metrics like precision,~P = TP/(TP+FP) and recall,~R = TP/(TP+FN). 
For a multi-class classification, one can compute the precision and recall for each class and average these values over the number of classes. This method is called macro-averaging and the value obtained is the macro-average precision. The results from the classification network are presented in figure~\ref{fig:classify_conf_matrix}, as normalized confusion matrices. We begin our analysis on GW simulations that represent clean signals, i.e.,~they do not incorporate any noise. The results from the classification neural network in the case of clean signals are shown in the top left corner of figure~\ref{fig:classify_conf_matrix}, and is as anticipated, highly accurate. The macro-average precision in this case is 1.0.
We further analyze the signals that incorporate noise, where we increase the noise amplitude in consecutive steps. 
The confusion matrix for the network classifier in the case of SNR=30 is depicted in the top right corner of figure~\ref{fig:classify_conf_matrix}. The macro-average precision in this case also reaches a high value of up to $\sim$1.0. The classification network takes longer to learn when the SNR is reduced to 25. We observe an increase in misclassifications in the confusion matrix~(figure~\ref{fig:classify_conf_matrix}, bottom left). A macro-average precision of 0.99 is attained in this case, however only after 60 epochs of training. 
Here, we observe that the performance does not decrease significantly with pSNR=25, despite requiring more epochs to reach a high accuracy. Similarly, in the case of SNR=20, we observe that training the network models for a larger number of epochs yields a higher accuracy (after 100 epochs of training). The corresponding confusion matrix is depicted in the bottom right corner of figure~\ref{fig:classify_conf_matrix}, and yields a macro-average precision value of 0.96.

%%%%%%%%%%%%%%%%%%%%%%%%%%%%%%%%%%%%%%%%%%%%%%%%%%%%%%%%
%%%%%%%%%%%%%%%%%%%%%%%%%%%%%%%%%%%%%%%%%%%%%%%%%%%%%

\section{Regression of chirp mass and combined tidal deformability} \label{sec:Regression}
We discuss the application of DL for parameter estimation from GW signals of BNS mergers in this section. The estimation of the binary masses in BNS mergers has already been studied using deep learning methods~\cite{Krastev:2020skk}. 
However, we show in this study that DL methods can also be used to extract the chirp mass,~$\mathcal{M}$, and combined tidal deformability,~$\Tilde{\Lambda}$, from GW signals of BNS mergers. 
We develop a regression network to estimate $\mathcal{M}$ and $\Tilde{\Lambda}$, rather than the component masses,~$\{m_1,m_2\}$ and individual tidal deformabilities,~$\{\Lambda_1,\Lambda_2\}$. In order to disentangle the individual masses from the chirp mass, one requires an estimate of the mass ratio~($q$) which is difficult to obtain due to the degeneracy with the aligned spin values in higher order PN expansion terms~\cite{Cutler:1994ys}. 
Similarly, in addition to $\Tilde{\Lambda}$, an estimate of the corrections to the tidal parameter~($\delta\Tilde{\Lambda}$) is required to disentangle the individual tidal deformabilities. This however appears only at 6PN order~\cite{Damour:2012yf, Chatziioannou:2020pqz, Wade:2014vqa}, making $\Tilde{\Lambda}$ an easier choice for regression~\cite{Flanagan:2007ix, Hinderer:2009ca, Chatziioannou:2020pqz}. 
In subsection~\ref{subsec:eos_constraints}, we illustrate that the estimation of chirp mass and combined tidal deformability can provide good constraints on the underlying EoS. The description of the DL models and results follow in subsections~\ref{subsec:dl_regression} and \ref{subsec:results_regression}, respectively.

\subsection{EoS Constraints from  $\mathcal{M}-\Tilde{\Lambda}$ relation} \label{subsec:eos_constraints}
\begin{figure}[!htbp]
    \centering
    \includegraphics{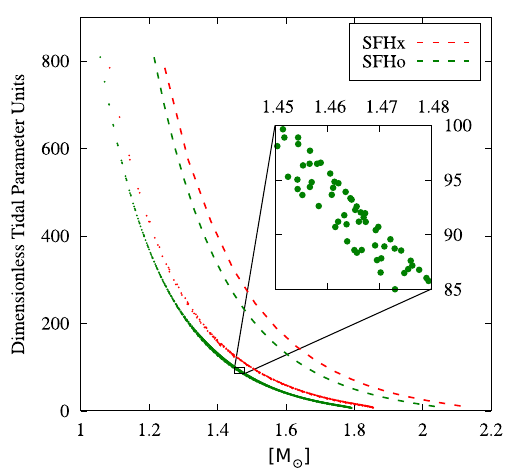}
    \caption{$\text{M}-\Lambda$ and $\mathcal{M}-\Tilde{\Lambda}$ curves of SFHx~(red) and SFHo~(green) EoSs. The dashed lines depict the $\text{M}-\Lambda$ curve. The solid points are $\mathcal{M}-\Tilde{\Lambda}$ values, obtained using several plausible combinations of component masses and tidal deformabilities of a particular EoS. The inset image displays a finer resolution of the $\mathcal{M}-\Tilde{\Lambda}$ space within a specific range. The area spanned by the $\mathcal{M}-\Tilde{\Lambda}$ points of the SFHo EoS in the zoomed-in region is not broader than $\pm5$ dimensionless units for $\pm0.01M_{\odot}$.}
    \label{fig:td_cm_cl_sfhox}
\end{figure}

Due to the systematic uncertainties of the component masses and individual tidal deformabilities when provided with $\mathcal{M}$ and $\Tilde{\Lambda}$ alone, it is not possible to obtain a specific mass-tidal deformability~($\text{M}-\Lambda$) relation. Therefore, it follows that the corresponding EoS remains ambiguous, too. 
However, the inferred $\mathcal{M}-\Tilde{\Lambda}$ relations can still provide ample constraints on the dense matter EoS. This is demonstrated in figure~\ref{fig:td_cm_cl_sfhox}, where the $\mathcal{M}-\Tilde{\Lambda}$ relations are shown as points and the $\text{M}-\Lambda$ relations as dashed lines~\footnote{Note: Also compare to Fig.~3 of~\cite{Chatziioannou:2020pqz}.}. The $\mathcal{M}-\Tilde{\Lambda}$ relations are obtained by using all possible combinations of $(m_1,\Lambda_1)$ and $(m_2,\Lambda_2)$ from a particular EoS, such that $m_1 \gtrsim m_2$. The figure depicts the relations of two EoSs, SFHo and SFHx~\cite{Steiner:2012rk}, given in green and red, respectively. 
There is a clear distinction between the $\mathcal{M}-\Tilde{\Lambda}$ points from the two EoSs. 
The inset image shows a few points from the SFHo EoS on the $\mathcal{M}-\Tilde{\Lambda}$ space. 
It can be observed that within the focused range, changing the chirp mass by $\mathcal{M}\pm$ 0.01~$M_{\odot}$ only leads to sizeable variations, $\Tilde{\Lambda}\pm$5. This implies that the $\mathcal{M}-\Tilde{\Lambda}$ relations obtained from the analysis of GW signals of BNS mergers can add further constraints on the EoS. It is also known that $\Tilde{\Lambda}$ is almost insensitive to the mass ratio~\cite{Chatziioannou:2020pqz}. Therefore, given $\mathcal{M}$, and narrow bounds for $\Tilde{\Lambda}$, several EoSs can be ruled out. However, a large number of observations are required to obtain sufficient points on the $\mathcal{M}-\Tilde{\Lambda}$ space. 
The uncertainties on the combined tidal deformability from GW170817 and AT2017gfo,~($197\leq\Tilde{\Lambda}\leq720$)~\cite{LIGOScientific:2018mvr, Coughlin:2018miv}, despite being large, can already rule out extremely stiff EoSs~\cite{Soma:2019utv}. Hence, with an increase in BNS merger detections with higher confidence, the extraction of combined tidal deformability plays a significant role in constraining the dense matter EoS. 

%%%%%%%%%%%%%%%%%%%%%%%%%%%%%%%%%%%%%%%%%%%%%%%%
%%%%%%%%%%%%%%%%%%%%%%%%%%%%%%%%%%%%%%%%%%%%%%%%

\subsection{Regression: Network Architecture} \label{subsec:dl_regression}
We use BNS inspiral-merger-ringdown waveforms generated using the methods described in the previous subsection for training and testing the regression network.  
We consider the cases with clean signals as well as those with noise. The regression network in each case is a convolutional neural network~(CNN), i.e., structured with several 1D convolutional layers. 

\subsubsection{Independent parameter regression (without noise)}
\begin{figure*}[htb!]
    \centering
    \includegraphics[width=0.92\textwidth]{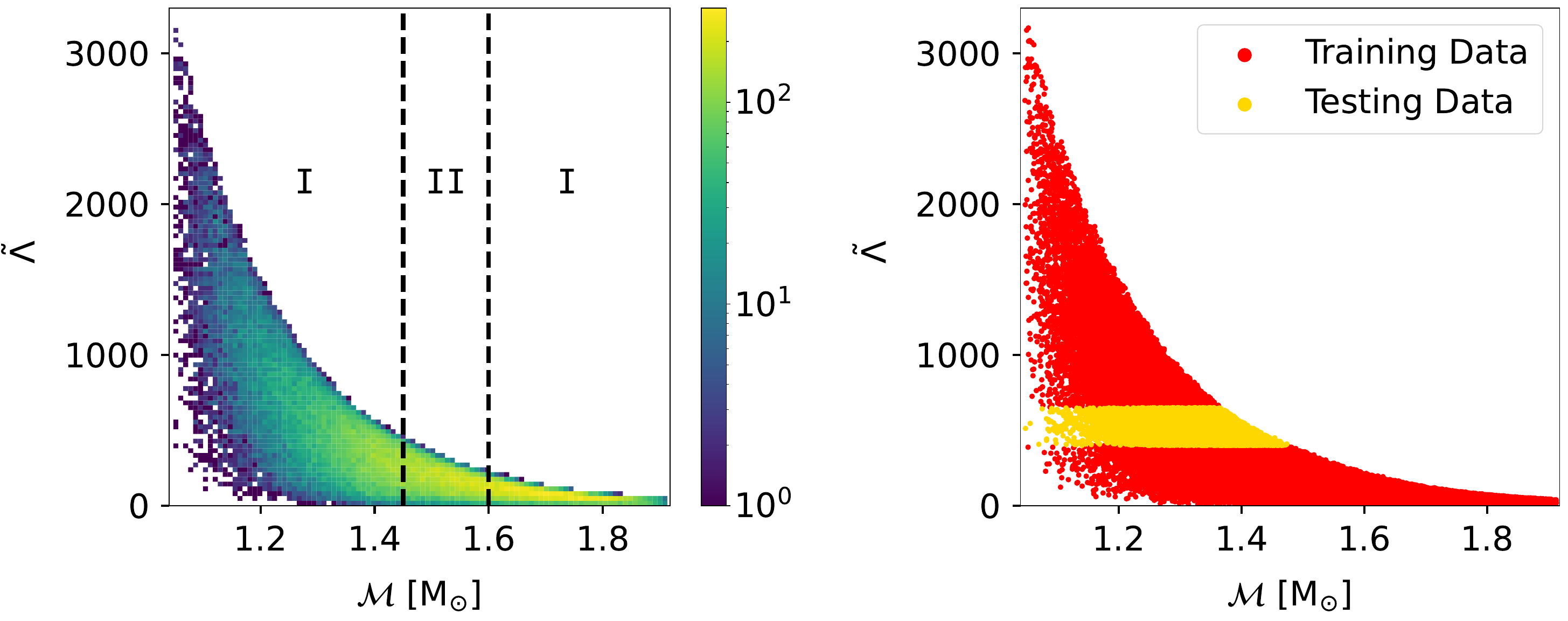}
    \caption{ Choice of training and testing data for independent parameter regression. $\textit{Left}$: Density plot of chirp mass and tidal deformability~($\mathcal{M}-\Tilde{\Lambda}$) values for all simulated GW events. Regions $\mathtt{I}$ contain training samples for the $\mathcal{M}$-regression. Region $\mathtt{II}$ depicts the range of the testing samples. The total number of samples amounts to 75,000. Note that this density distribution of $\mathcal{M}-\Tilde{\Lambda}$ is the same in the right panel, as well as in all subplots of figure~\ref{fig:regress_sim_test_train}. {\it Right:} Data distribution used for $\Tilde{\Lambda}$-regression. Red and yellow points denote the training and testing data, respectively. }
    \label{fig:regress_sep_test_train}
\end{figure*}

For the primitive tests, we focus on estimating only one of the two parameters, i.e., regressing the chirp mass,~$\mathcal{M}$, and the combined tidal parameter,~$\Tilde{\Lambda}$ independently. Just like the two-channel input to the classification network, we utilize both the amplitude,~$A$, and phase,~$\phi$, as input to the regression network as well. The primitive network for the regression task is designed for fairly sparse input data. Here, we use $f_{\text{low}}$~=~128~Hz, $f_{\text{high}}$~=~1024~Hz and $\Delta f$~=~1/16. Therefore, the input shape to the regression network is (14336,~2). The network consists of five 1D~convolutional layers, two max-pooling layers for down-sampling~(pool size~=~2) and two dense layers. The first pooling layer follows the second convolutional layer, and the second pooling layer follows the fourth convolution layer. 
The convolutional layers have 16, 16, 32, 32 and 16 filters in that specific order. Kernels sizes of 32, 32, 24, 24 and 16 are used in this respective order. The kernel weights are initialized from the {\it{He}}~Normal distribution~\cite{he2015delving}. A stride of 3 is applied to the first four convolutions, and a stride of 2 to the last convolutional layer. We apply the `ReLU' activation function to all the convolutional layers. The last layer uses the `sigmoid' activation. We use the mean squared error~(MSE) for the loss function, and an Adam optimizer~\cite{2014arXiv1412.6980K} with a learning rate of 0.0001. The model accuracy is measured by the coefficient of determination, $\mathcal{R}^2$, which is defined as $ 1 - {\sum_i(y_i - \hat{y_i})^2}/({\sum_{i}(y_i - \bar{y})^2 + \delta})$. Here, $y_i$ is the true value of the $i^{\text{th}}$ chirp mass~($\mathcal{M}$) or $i^{\text{th}}$ tidal parameter~($\Tilde{\Lambda}$), $\hat{y_i}$ is the corresponding prediction from the regression network, and $\bar{y}$ is the mean of the true $\mathcal{M}$ or $\Tilde{\Lambda}$ values. We set $\delta=10^{-7}$ to avoid undefined values when encountered with a division by zero. We normalize the chirp mass and combined tidal deformability such that they lie in the range~[0,1]. Therefore, the labels, $\mathcal{M},\Tilde{\Lambda}$, are normalized as,
\begin{equation}
    y_{\text{norm}} = (y - y_{\text{min}})/(y_{\text{max}}-y_{\text{min}}),
    \label{eq:norm}
\end{equation}
where, $y=\{\mathcal{M},\Tilde{\Lambda}\}$. The two input channels, the Fourier amplitude and the phase, are normalized using the same method as for the classification network. We generate the complete data set using the parameters $\{m_1, m_2, \Lambda_1, \Lambda_2\}$. These parameter values, when represented in terms of $\mathcal{M}$ and $\Tilde{\Lambda}$, span the region shown by the points in figure~\ref{fig:regress_sep_test_train}. The left panel represents a density plot of the parameters chosen for all simulations. The total number amounts to 75,000 simulated events. Training and testing data for the $\mathcal{M}$-regression were chosen as depicted by regions $\mathtt{I}$ and $\mathtt{II}$, respectively. The right panel illustrates the training and testing data used for regressing $\Tilde{\Lambda}$ as red and yellow points, respectively. The distribution of the events across both the parameters, $\mathcal{M}$ and $\Tilde{\Lambda}$, is the same as on the left panel. Each point on the right panel of the figure represents the source parameters of one GW event. This kind of segregation of training and testing data was chosen to ensure that the network can extrapolate to regions not covered during the training process. For the $\mathcal{M}$-regression, we use GW signals generated with $\mathcal{M}$ in the range $\mathcal{M}\in$~[1.45,~1.6]~$M_{\odot}$ as test data. This amounts to 54,776 training samples and 20,224 test samples for the $\mathcal{M}$-regression. For the $\Tilde{\Lambda}$-regression, we use GW signals generated with $\Tilde{\Lambda}\in$~[400,~650] as test data. The training and test samples for $\Tilde{\Lambda}$-regression then sum up to 64,532 and 10,468 respectively. A batch size of 16 is used for both $\mathcal{M}$- and $\Tilde{\Lambda}$-regression. The networks are trained for 50 epochs in both cases. \\
%%%%%%%%%%%%%%%%%%%%%%%%%%%%%%%%%%%%%%%%%%%%%%%%%%%%%%
%%%%%%%%%%%%%%%%%%%%%%%%%%%%%%%%%%%%%%%%%%%%%%%%%%%%%%%%

\subsubsection{Simultaneous parameter regression (without noise)}
We present three distinct test cases in this section. The major differences arise either due to data-segregation or hyperparameter choices. \\

\begin{figure*}[htb!]
    \centering
    \includegraphics[width=1.\textwidth]{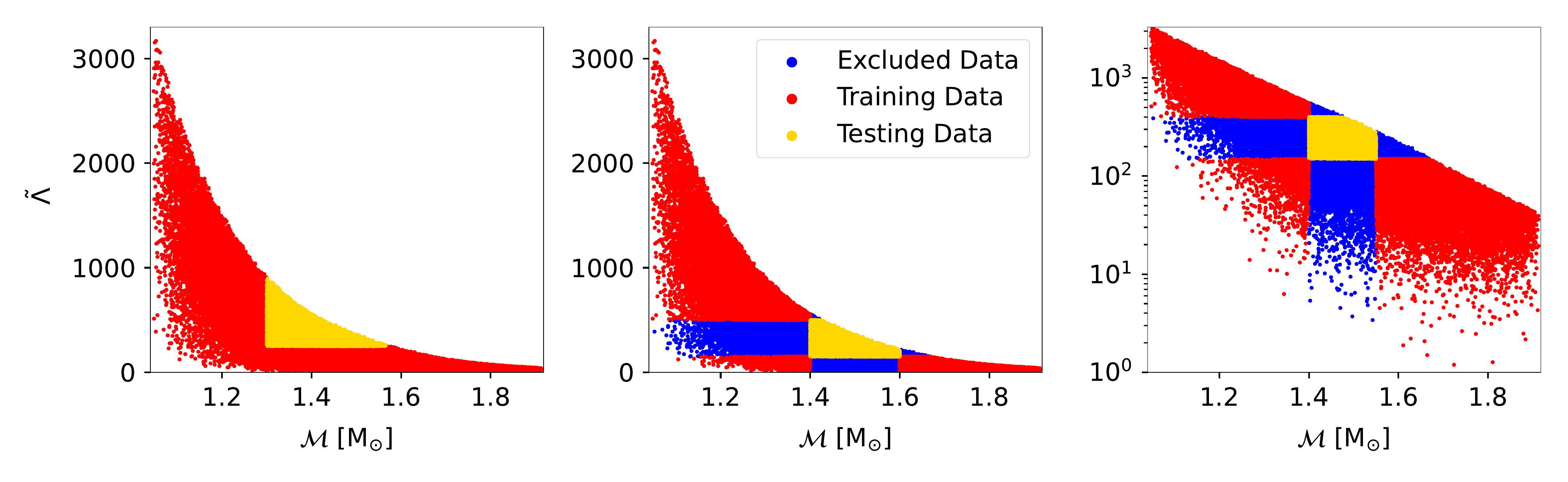}
    \caption{{\it Left}: Training~(red) and test~(yellow) data chosen for Case I of the simultaneous regression of $\mathcal{M}$ and $\Tilde{\Lambda}$. {\it Centre}: Same as the left figure for Case II of the simultaneous parameter regression. In contrast to the subfigure on the left, the blue data points in this figure are excluded in the training process in order to test the robustness of the network. {\it Right}: Same as centre figure, for Case III of the simultaneous parameter regression. Note that we depict the tidal parameter as a $\log$-scale here as the data labels for the tidal parameter are substituted for $\log(\Tilde{\Lambda})$ in this case.  }
    \label{fig:regress_sim_test_train}
\end{figure*}

Case I: Using the same network structure described for independent parameter regression, with a slight modification to the last dense layers, we train the weights of the regression network to output both $\mathcal{M}$ and $\Tilde{\Lambda}$, simultaneously. The network structure and the layer dimensions can be found in table~\ref{reg_cnn}. The network is no longer a sequential model, as the last dense layer~(layer index~10) branches into two dense layers~(10a and 10b), one to output the chirp mass and the other to output the combined tidal deformability. 
The same values of $f_{\text{low}}$~=~128~Hz, $f_{\text{high}}$~=~1024~Hz and $\Delta f$~=~1/16 are applied here. The loss functions, activations and normalization techniques of the input and labels remain the same as in the previous case. We test the performance of the network on a simple variation of data-segregation. We use the training and testing data as shown in the left panel of figure~\ref{fig:regress_sim_test_train} for this case. The distribution of events is the same as shown in the density plot of figure~\ref{fig:regress_sep_test_train}. The training and testing data are represented by red and yellow, respectively. This choice of a test set, where $\mathcal{M}\in$~[1.3,~1.57]~$M_{\odot}$ and $\Tilde{\Lambda}\in$~[250,~895], amounts to 19582 test samples. The remaining data, i.e. 55418 samples are used for training the network. 
The network is trained for 100 epochs with a batch size of 16. The results are presented in subsection,~\ref{subsec:results_regression}. \\
%%%%%%%%%%%%%%%%%%%%%%%%%%%%%%%%%%%%%%%%%%%%%%%%%%%%%%
\begin{table}[ht!]
\caption{Example of a model architecture used for the simultaneous regression of chirp mass~$\mathcal{M}$ and combined tidal deformability~$\Tilde{\Lambda}$ (Case I and Case II). The network comprises of 66,866 trainable parameters.}
\centering
\def\arraystretch{1.4}
\setlength\tabcolsep{12pt}
\begin{tabular}{@{}ccc@{}}
\hline\hline
Layer Index & Layer & Dimension\\
\hline%
1 & Input & 14336 x 2\\
2 & Convolution 1D & 4769 x 16\\
3 & Convolution 1D & 1580 x 16\\
4 & Max Pooling & 790 x 16\\
5 & Convolution 1D & 256 x 32\\
6 & Convolution 1D & 78 x 32\\
7 & Max Pooling & 39 x 32\\
8 & Convolution 1D & 12 x 16\\
- & Reshape & 192\\
9 & Dense Layer & 64\\
10a & Dense Layer & 1\\
10b & Dense Layer & 1\\
%- & Output & 1\\
\hline\hline
\end{tabular}
\label{reg_cnn}
\end{table}
%%%%%%%%%%%%%%%%%%%%%%%%%%%%%%%%%%%%%%%%%%%%%%%%%%%%%%%%

Case II: In order to successfully increase the robustness of the regression network, we generalize the segregation of training and testing data from the previous case of independent parameter regression. For this, we provide the network with training and testing data as depicted in the centre panel of figure~\ref{fig:regress_sim_test_train}. The blue data points are excluded from the training set to mask an entire range of both parameters to the regression network. The testing set includes only the points from the masked ranges marked in yellow. Therefore, the blue points are redundant to our analysis. The red and yellow data points represent the training and testing data respectively. We apply a test-train data-segregation where GW signals generated from a range of $\mathcal{M}\in$~[1.4,~1.6]~$M_{\odot}$ and $\Tilde{\Lambda}\in$~[150,~500] are used as test data. This adds up to 18,718 samples in the testing data. The training data includes simulated GW signals with source parameters that do not lie in the indicated ranges of the testing set, and therefore comprises of 33,686 waveform samples. With this kind of data-segregation, we train the network for 200 epochs. Longer training is required with the chosen ratio of the testing to training samples~($\sim~$0.56) in this case.\\ 

Case III: We present a third scenario where we tune the hyperparameters of the network. Due to the non-linear dependence of $\Tilde{\Lambda}$ on $\mathcal{M}$, in addition to the previously trained regression networks, we assess the simultaneous regression of the parameters by using the tidal deformability labels as $\log(\Tilde{\Lambda})$. Furthermore, we train the network weights on a different form of input data in this case. The frequency-domain waveforms are represented in terms of real and imaginary parts. We use a fine representation of the waveforms in this case. The waveforms include signals within the frequency range~[23,~2048]~Hz, with $\Delta f$~=~1/128.
This implies an input dimension of (259454,~2).
The real and imaginary parts of the frequency domain waveforms are normalized to have an integral sum of unit value each. Both the channels are then rescaled to lie within the range~[0,~1]. Normalization of the tidal parameter is similar to Eq.~\ref{eq:norm}, but in this case, $y = \log(\Tilde{\Lambda} + 1)$. The max-pooling layers are replaced with average-pooling layers in this network. We further introduce an L2 regularizer to the convolutional kernels. The dense layer~(index 9 in table~\ref{reg_cnn}) is modified to have 128 nodes. The activation function of this particular dense layer is updated to `Scaled Exponential Linear Unit~(SELU)'. Additionally, the activation functions acting on the rest of the convolutional layers are updated from `ReLU' to `Exponential Linear Unit~(ELU)'. Furthermore, due to the unusually large ratio of testing to training samples in Case~II, we reduce the range of the testing samples in this case. The training and testing data is depicted in the right panel of figure~\ref{fig:regress_sim_test_train}. We use a $\log$-scale for $\Tilde{\Lambda}$, as in this case the network is trained on $\log(\Tilde{\Lambda})$. The trained network is tested on $\mathcal{M}\in$~[1.4,~1.55]~$M_{\odot}$ and $\Tilde{\Lambda}\in$~[150,~400]. Applying the same techniques for test-train data-segregation results in 14,808 testing samples and 41,588 training samples. This way, we improve the ratio of the testing-training samples to $\sim$~0.36.

%%%%%%%%%%%%%%%%%%%%%%%%%%%%%%%%%%%%%%%%%%%%%%%%%%%%%%
%%%%%%%%%%%%%%%%%%%%%%%%%%%%%%%%%%%%%%%%%%%%%%%%
%%%%%%%

\subsubsection{Simultaneous parameter regression (with noise)}
We further extend the work to include noise. For this, we use gravitational waveforms that are generated with noise as described earlier, and whiten the simulated signals such that the power of noise contributes equally at varying frequency intervals. The data is input to the network in terms of real and imaginary parts of the frequency domain waveform. 
The waveforms are modeled with frequency range~[23,~2048]~Hz, with $\Delta f$~=~1/128, implying an input dimension of (259454,~2). The same activation functions as in Case~III for the simultaneous regression are applied. We alter the activation function of the dense layer that outputs $\Tilde{\Lambda}$, from `Sigmoid' to a `tangent hyberbolic~(tanh)' function. Due to the steep gradients, the learning step-size is expected to increase in the regression of tidal deformability. The first step in normalization of the tidal parameter is similar to Eq.~\ref{eq:norm}, but in this case, $y = \log(\Tilde{\Lambda} + 1)$. The second step involves a rescaling of $y_{\text{norm}}$ from [0,~1] to [-1,~1] as a requirement for the 'tanh' activation function. Therefore, we further rescale the output from Eq.~\ref{eq:norm} as $y_{\text{norm}}^{'}=2(y_{\text{norm}})-1$. The training and testing data are chosen randomly in this case. We generate a total of 48,000 waveforms with noise, where 36,000 samples are randomly selected for training, and the remaining 12,000 samples for testing the network. We train the network in batches of 50 samples for 80 epochs. 

\subsection{Regression: Results} \label{subsec:results_regression}
In this subsection, we discuss the results obtained from the regression networks, with and without noise, as detailed in the subsection~\ref{subsec:dl_regression}. \\
\subsubsection{Independent parameter regression (without noise)}
\begin{figure*}[htbp]
    \centering
    \includegraphics[width=0.97\textwidth]{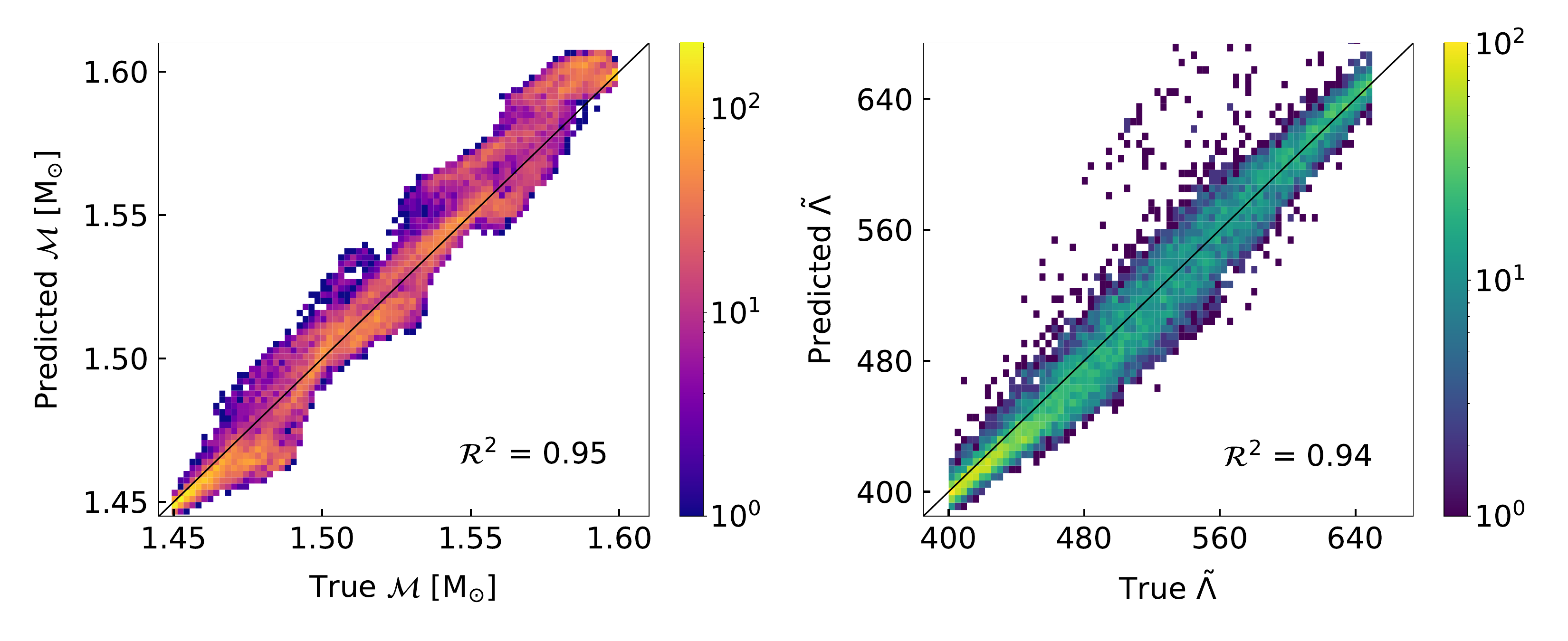}
    \caption{Predicted parameters plotted against true parameters for the case of independent parameter regression (without noise) for $\mathcal{M}$~(left panel) and $\Tilde{\Lambda}$~(right panel). Both subfigures are density plots. The distribution of the points becomes sparse as one moves away from the diagonal line. The total number of testing samples used for the independent $\mathcal{M}$- and $\Tilde{\Lambda}$-regression amount to 20224 and 10468, respectively.}
    \label{fig:reg_cm_cl_sep_nonoise}
\end{figure*}

Two individual networks are trained to estimate $\mathcal{M}$ and $\Tilde{\Lambda}$, independently. This implies that while training the network to estimate the chirp mass, no explicit information on the tidal parameter is provided to the network. Any information to the $\mathcal{M}$-regression network regarding the tidal parameter is only implicit through the waveforms and {\it vice~versa}. The results for $\mathcal{M}$-regression and $\Tilde{\Lambda}$-regression from training two independent networks are presented in figure~\ref{fig:reg_cm_cl_sep_nonoise} in the left and right panels, respectively. The sub-figures depict the predicted parameters against the true parameters. The plots are represented as 2D-histograms, where the shading provides information on the density of points. The diagonal black solid line is the function, $y=x$, characterizing a network with 100\% accuracy. The figure on the left represents the test events used in the $\mathcal{M}$-regression. The right panel depicts the events used to test the trained network on the $\Tilde{\Lambda}$-regression. As seen from both the left and right panel of figure~\ref{fig:reg_cm_cl_sep_nonoise}, a high density of chirp mass and combined tidal deformability values lie around the black solid line. The distribution of the data decreases as one moves away from the diagonal line. The simple independent parameter regression yields $\mathcal{R}^2$ values of 0.95 and 0.94 for the $\mathcal{M}$ and $\Tilde{\Lambda}$ respectively. 

%%%%%%%%%%%%%%%%%%%%%%%%%%%%%%%%%%%%%%%%%%%%%%%%
%%%%%%%%%%%%%%%%%%%%%%%%%%%%%%%%%%%%%%%%%%%%%%%%

\subsubsection{Simultaneous parameter regression (without noise)}
The results for the Cases I, II and III are presented below.\\

\begin{figure*}[htbp]
    \centering
    \includegraphics[width=0.97\textwidth]{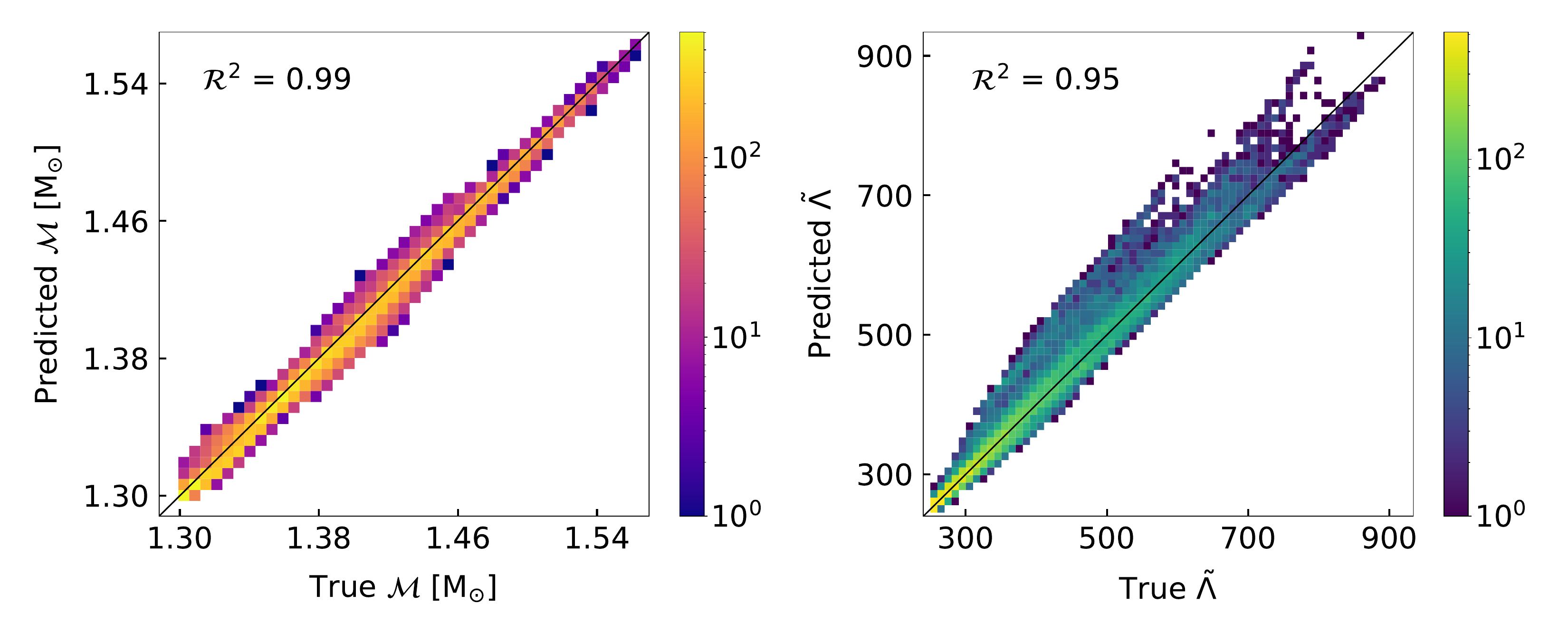}
    \caption{The predicted parameters plotted against the true parameters, shown as density plots in the case of simultaneous parameter regression (Case~I, without noise) for $\mathcal{M}$~(left panel) and $\Tilde{\Lambda}$~(right panel). Testing samples amount to 19582 in this case. An improvement in accuracy from figure~\ref{fig:reg_cm_cl_sep_nonoise} is observed here due to the choice of test-train data segregation (in this case, the network is not completely obscure to an entire range of parameters during the training process, see figure~\ref{fig:regress_sim_test_train}.) }
    \label{fig:reg_cm_cl_sim_nonoiseC1}
\end{figure*}

Case~I: The results of the simultaneous regression with training and testing data corresponding to left panel of figure~\ref{fig:regress_sim_test_train}, are presented in figure~\ref{fig:reg_cm_cl_sim_nonoiseC1}. Note that the color representation for $\mathcal{M}$ and $\Tilde{\Lambda}$ follows from figure~\ref{fig:reg_cm_cl_sep_nonoise}, and remains the same for all figures in section~\ref{subsec:results_regression}. However, in this case and all the ones to follow, both parameters are estimated simultaneously, i.e. from a single network. The trained network results in high accuracy values for both parameters, i.e. $\mathcal{R}^2$=~0.99 for the $\mathcal{M}$-regression and $\mathcal{R}^2$=~0.95 for the $\Tilde{\Lambda}$-regression. The high accuracy values for both parameters can be attributed to the fact that the network is not completely devoid of an entire range of $\mathcal{M}$ and $\Tilde{\Lambda}$ during the training process in this case.\\

\begin{figure*}[htbp]
    \centering
    \includegraphics[width=0.97\textwidth]{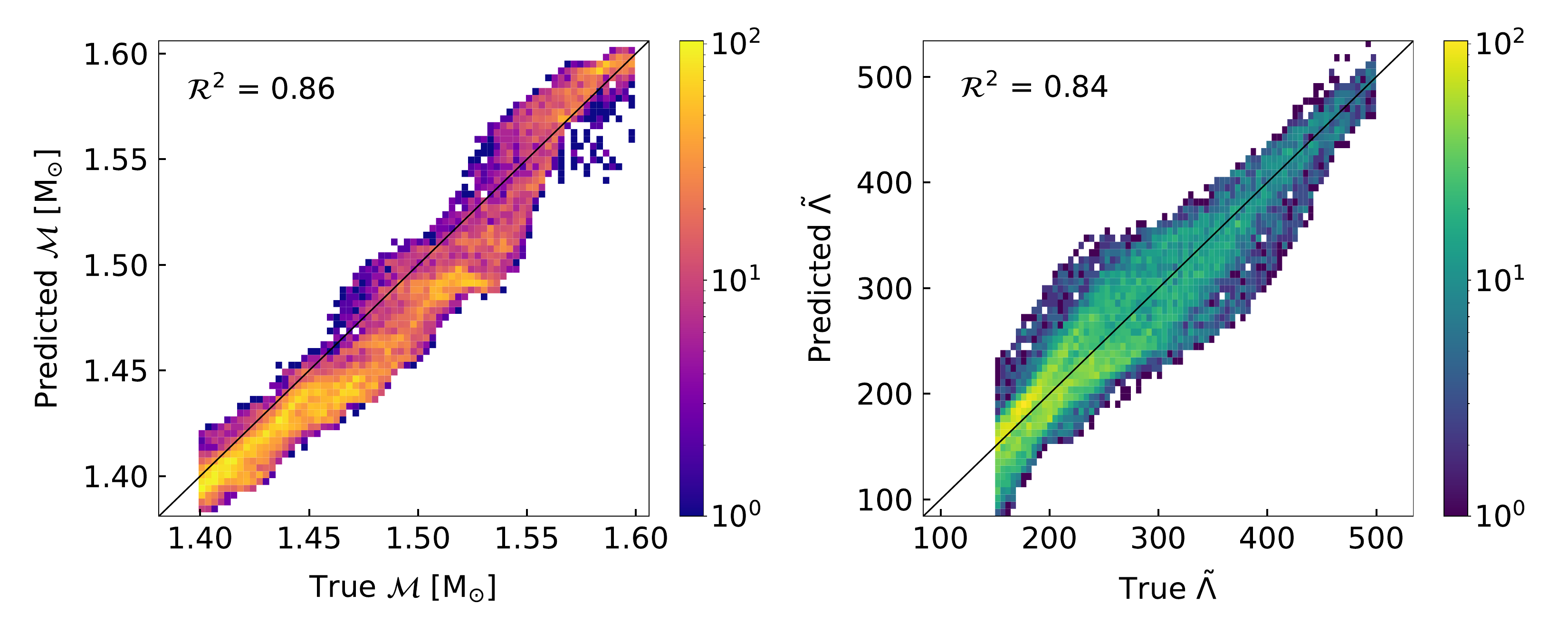}
    \caption{Same as figure~\ref{fig:reg_cm_cl_sim_nonoiseC1}, for Case II of the simultaneous parameter regression, with a total of 18718 testing samples. The network faces challenges when an entire parameter range is masked during the training process. Additionally, the $\sim$65\%-35\% splitting between the training and the testing data here, leads to considerably lower accuracies in this case.}
    \label{fig:reg_cm_cl_sim_nonoiseC2}
\end{figure*}

Case II: Eliminating an entire range of $\mathcal{M}$ and $\Tilde{\Lambda}$ in the training process offers new challenges to the network. The results depicting the true and predicted parameters in this case are presented in figure~\ref{fig:reg_cm_cl_sim_nonoiseC2}. This entails a test range of $\mathcal{M} \in$~[1.4,~1.6] and $\Tilde{\Lambda} \in$~[150,~500]. The network outputs $\mathcal{R}^2$=~0.86 for the $\mathcal{M}$-regression and $\mathcal{R}^2$=~0.84 for the $\Tilde{\Lambda}$-regression. The $\mathcal{R}^2$ values decrease for both parameters due to the challenges the network faces when it is blind to a large range of parameters. The small number of training samples when compared to the testing samples (or a large value of test-to-train samples) also plays a role in the reduction of $\mathcal{R}^2$.\\

\begin{figure*}[tbp]
    \centering
    \includegraphics[width=0.97\textwidth]{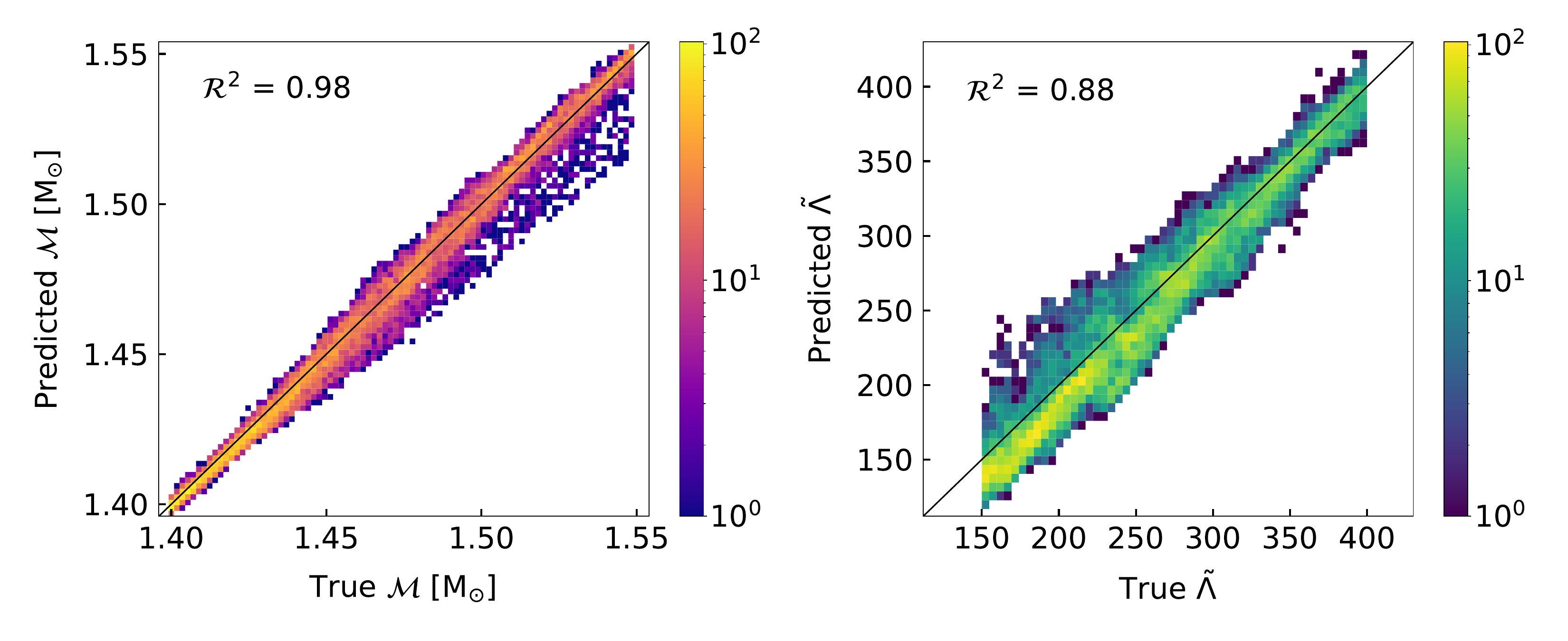}
    \caption{Same as figure~\ref{fig:reg_cm_cl_sim_nonoiseC1}, for Case III of the simultaneous parameter regression, with 14808 testing samples. The use of different scaling for combined tidal deformability, along with the updates to certain hyperparameters in the training process, leads to an improvement in accuracies in this case~(when compared to Case II, figure~\ref{fig:reg_cm_cl_sim_nonoiseC2}).}
    \label{fig:reg_cm_cl_sim_nonoiseC3}
\end{figure*}

Case III: The results of the true and predicted parameters in this case are presented in figure~\ref{fig:reg_cm_cl_sim_nonoiseC3}. The range of both parameters that the network is uninformed about, is reduced to $\mathcal{M} \in$~[1.4,~1.55] and $\Tilde{\Lambda} \in$~[150,~400]. Apart from the several modifications applied to the hyperparameters in the network in this case, the update from  $\Tilde{\Lambda}$ labels to $\log(\Tilde{\Lambda})$ results in an improvement of the $\mathcal{R}^2$ values of both parameters. We obtain $\mathcal{R}^2$=~0.98 for the $\mathcal{M}$-regression and $\mathcal{R}^2$=~0.88 for the $\Tilde{\Lambda}$-regression. The ratio of the test-train samples also helps producing more reliable results. 

%%%%%%%%%%%%%%%%%%%%%%%%%%%%%%%%%%%%%%%%%%%%%%%%
%%%%%%%%%%%%%%%%%%%%%%%%%%%%%%%%%%%%%%%%%%%%%%%%
%%%%%%%%%%%%%%%%%%%%%%%%%%%%%%%%%%%%%%%%%%%%%%%%
%%%%%%%%%%%%%%%%%%%%%%%%%%%%%%%%%%%%%%%%%%%%%%%%
\subsubsection{Simultaneous parameter regression (with noise)}
\begin{figure*}[tbp]
    \centering
    \includegraphics[width=0.97\textwidth]{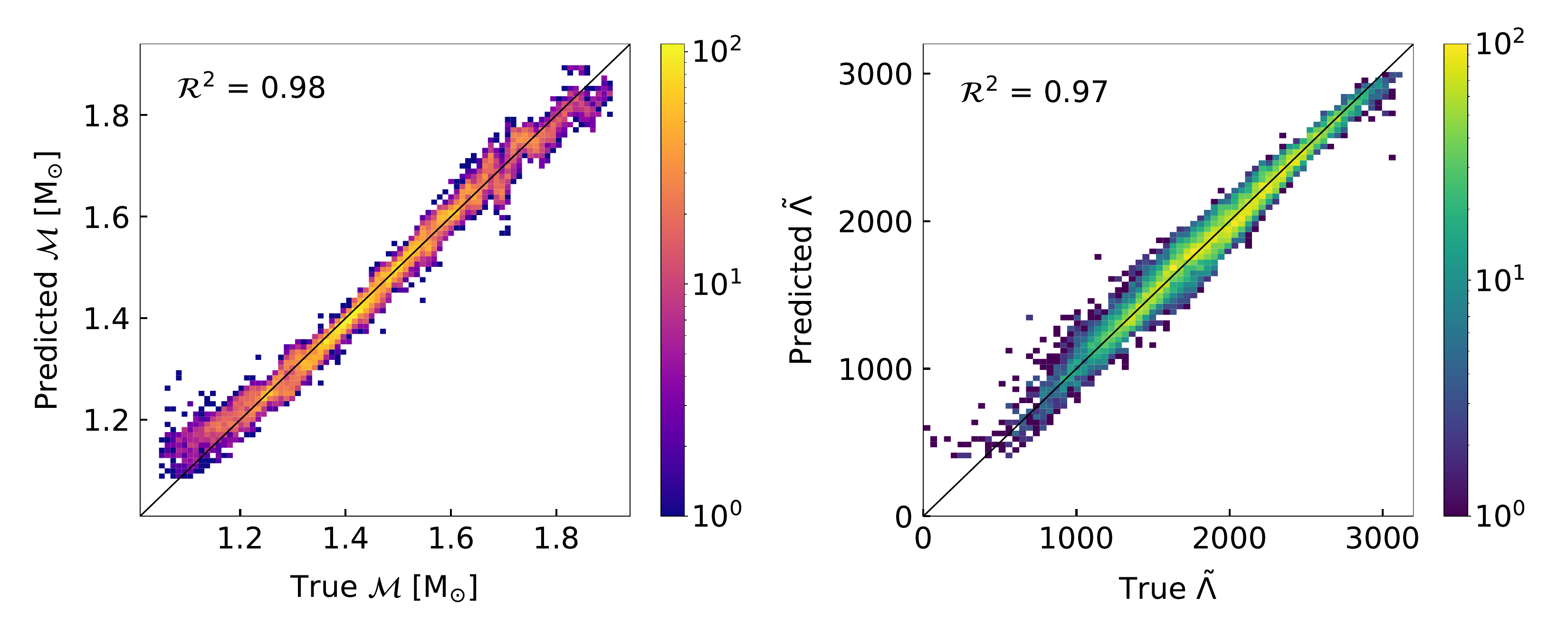}
    \caption{Density plots of the predicted parameters plotted against true parameters in the case of simultaneous parameter regression (with noise) for $\mathcal{M}$~(left panel) and $\Tilde{\Lambda}$~(right panel). 12000 samples were used for testing. In this case, we use a random segregation of testing-training data, uniformly spread across the entire range. The network therefore does not face challenges when dealing with testing data~(as observed in figures~\ref{fig:reg_cm_cl_sim_nonoiseC2} and \ref{fig:reg_cm_cl_sim_nonoiseC3}, i.e., Case II and III of simultaneous parameter regression). Hence, we obtain high accuracies for both parameters simultaneously in this case, despite the inclusion of noise.}
    \label{fig:reg_cm_cl_sim_noise}
\end{figure*}

The results of the regression network trained on signals that incorporate noise are shown in figure~\ref{fig:reg_cm_cl_sim_noise}. The choice of random values of $\mathcal{M}$ and $\log(\Tilde{\Lambda})$ allows for an easier prediction of parameter values. The inclusion of the complete range of parameters without withholding data from a specific plausible range of $\mathcal{M}$ and $\Tilde{\Lambda}$, produces more accurate results, as the network has the capacity to interpolate between the points in the training data. The network outputs $\mathcal{R}^2$ values of 0.98 and 0.97 for the prediction of $\mathcal{M}$ and $\Tilde{\Lambda}$, respectively.

%%%%%%%%%%%%%%%%%%%%%%%%%%%%%%%%%%%%%%%%%%%%%%%%
%%%%%%%%%%%%%%%%%%%%%%%%%%%%%%%%%%%%%%%%%%%%%%%%

\section{Conclusions and discussions}\label{sec:discuss}
This work demonstrates the capability and performance of neural networks for the analysis of simulated gravitational data. We perform two distinct tasks: (1) classification of GW signals from binary black hole mergers, from binary neutron star mergers and signals which contain only noise, and (2) regression of two source parameters from simulated GW signals of binary neutron star mergers, namely the chirp mass~($\mathcal{M}$) and the combined tidal deformability~($\Tilde{\Lambda}$). 
The trained classification network is able to detect the whitened signals embedded in detector noise with a macro-averaged precision of 0.99 for SNR=25, and a macro-averaged precision of 0.96 for SNR=20. We additionally highlight the regression of $\mathcal{M}$ and $\Tilde{\Lambda}$ from BNS merger signals. We test the independent and simultaneous parameter estimation on simulated GW signals, starting with clean signals. For the best case of simultaneous regression, the network achieves $\mathcal{R}^2$ values of 0.99 and 0.95 for $\mathcal{M}$ and $\Tilde{\Lambda}$, respectively (see figure~\ref{fig:reg_cm_cl_sim_nonoiseC1}). We further use signals that are initially injected into aLIGO's colored detector noise, whitened before analysis. The regression network utilizes random values of training and testing samples in this case, as opposed to specific ranges for testing samples in the previous case. This permits an undemanding interpolation between training samples, thus resulting in high $\mathcal{R}^2$ values, 0.98 and 0.97, for the test samples of $\mathcal{M}$ and $\Tilde{\Lambda}$, respectively. The prediction of the tidal parameter is crucial for constraining the underlying NS EoS. The output of the regression network does not, however, disentangle the combined tidal deformability into the individual tidal deformabilities. Nor does the network unravel the individual component masses from the estimation of the chirp mass. This presents a realistic scenario where one has better estimates only on the $\mathcal{M}$ and $\Tilde{\Lambda}$ values to constrain the dense matter NS EoS. Note that, despite appearing only at 5PN order, it helps that the combined tidal deformability is of the order of a few 100 or more (for example, when compared to the mass ratio which is $<1$), making $\Tilde{\Lambda}$ more pronounced in a GW signal~\cite{Chatziioannou:2020pqz}.

With every subsequent run, an increase of the sensitivity of the GW detectors is expected. In the current run, O4, LIGO has the capacity to detect binary neutron star mergers at distances in the range of 130-150~Mpc. The target of the O4 run, however, is to reach 160-190~Mpc~\cite{KAGRA:2013rdx}. This implies an increased number of BNS merger detections. Therefore, an estimate of the tidal deformability from these events, could help to further constrain the dense matter EoS in NSs. Moreover, third generation telescopes, like the Einstein Telescope~(ET) and the Cosmic Explorer~(CE) are anticipated to have sensitivities that are an order of magnitude better than the current generation GW detectors. Together, the ET and CE are expected to detect over 100 BNS merger events per year with an oSNR value $>$30~\cite{Iacovelli:2022bbs}. 

The two tasks as described can then be combined to create a pipeline for a future analysis of GW signals. The detection of a BNS merger by the classification network can be forwarded to the regression network for parameter estimation, thereby creating a complete pipeline for GW analysis. 
The networks designed in this work are elementary. They can be further developed to incorporate additional parameters of the binary coalescing system like spins, inclinations and distance. Furthermore, this work is based on a specific waveform approximant, i.e. `IMRPhenomPv2\_NRTidalv2' for BNS merger simulations. This could introduce a model dependence in the trained network. Full general relativistic magnetohydrodynamic~(GRMHD) simulations of binary neutron star mergers in 3D are considerably more accurate for modelling waveforms. However, a single simulation increases the computational cost by several orders of magnitude. Therefore, full GRMHD simulations are a less favourable alternative, albeit being more accurate and model-independent.

This work depends on constraining the dense NS EoS using mass and tidal parameters. Unlike the one-to-one correspondence between the mass-radius~($\text{M}-R$) curve and the underlying EoS, the $\text{M}-\Lambda$ curves do not necessarily have a one-to-one relation with the EoS~\cite{Raithel:2022efm, Raithel:2022aee}. This has been demonstrated with different EoS models that undergo a first-order phase transition~(FOPT) at significantly different densities, but result in identical $\text{M}-\Lambda$ curves. The effect of this degeneracy can be scrutinized in future work. Independent radius measurements from NICER could also help break this degeneracy.

Another crucial open question which has not been considered in this study is the possibility of a phase transition from hadronic to deconfined quark matter at high densities and temperatures. Mergers of binary neutron stars can potentially harbour such extreme conditions, rendering possibilities to study these effects~\cite{Bauswein:2018bma, Most:2018eaw}. 
In spite of the use of a waveform approximant which models the inspiral, merger and ringdown components of a GW event, in the current study, we limit our analyses here, to a maximum GW frequency of 2048~Hz. This preference of a frequency range excludes information in the ringdown or post-merger phase of a BNS merger. Therefore, the present work shall be extended to include post-merger frequencies that reach as high as 4096~Hz. An extensive analysis of the post-merger GW signals of BNS mergers can help understand the existence of a possible FOPT. Studies of the post-merger GW analyses would also benefit from deep learning techniques with exponential boosts in the associated computational costs. We leave this for future work.

%\appendix
%\section{Some title}
%Please always give a title also for appendices.

\acknowledgments

The authors thank Sebastian Blacker, Manjunath Omana Kuttan, Dr.~Jan Steinheimer and Dr.~Anton Motornenko for useful discussions and helpful comments. 
The work is supported by (i) Deutscher Akademischer Austauschdienst - DAAD~(S.~Soma), (ii) F\&E Funding from GSI~(S.~Soma), (iii) the BMBF under the ErUM-Data project and the KISS consortium (05D23RI1) funded by the BMBF in the ErUM-Data action plan~(K.~Zhou), (iv) the AI grant of SAMSON AG, Frankfurt~(S.~Soma and K.~Zhou), and (v) the Walter Greiner Gesellschaft zur Förderung der physikalischen Grundlagenforschung e.V. through the Judah M. Eisenberg Laureatus Chair at Goethe Universität Frankfurt am Main~(H.~Stöcker). We also thank the NVIDIA Corporation for the donation of NVIDIA GPUs.

%\paragraph{Note added.} This is also a good position for notes added after the paper has been written.

% Bibliography

%% [A] Recommended: using JHEP.bst file
\bibliographystyle{JHEP}
\bibliography{biblio.bib}

%% or
%% [B] Manual formatting (see below)
%% (i) We suggest to always provide author, title and journal data or doi:
%% in short all the informations that clearly identify a document.
%% (ii) please avoid comments such as "For a review'', "For some examples",
%% "and references therein" or move them in the text. In general, please leave only references in the bibliography and move all
%% accessory text in footnotes.
%% (iii) Also, please have only one work for each \bibitem.

\end{document}